\documentclass[12pt]{article}
\usepackage{epsfig}
\newcommand{\beq}{\begin{equation}}
\newcommand{\eeq}{\end{equation}}
\newcommand{\beqs}{\begin{eqnarray}}
\newcommand{\eeqs}{\end{eqnarray}}
\newcommand{\tr}{\; \mathrm{tr}\;}

\newcommand{\intx}{\int d^d x\;}
\newcommand{\intxy}{\int d^d x \; d^d y\;}
\newcommand{\pb}{\bar\Psi}
\newcommand{\pbs}{\bar\psi}
\newcommand{\dirac}{\not\! \partial}
\newcommand{\qslash}{\not\! q}
\newcommand{\gur}{{\frac{c u_l}{(1+r_l)^2}}}

\begin{document}

\begin{titlepage}
\begin{flushleft}
       \hfill                      \texttt{hep-th/9607072}\\
       \hfill                       UUITP-14/96\\
\end{flushleft}
\vspace*{3mm}
\begin{center}
{\LARGE The Critical Exponents Of The Matrix Valued Gross-Neveu Model\\}
\vspace*{12mm}
{\large Gabriele Ferretti}\footnote{E-mail: \texttt{ ferretti@teorfys.uu.se}}
\\
\vspace*{4mm}
{\em Institutionen f\"{o}r teoretisk fysik \\
Box 803\\
S-751 08  Uppsala \\
Sweden \/}\\
\vspace*{10mm}
\end{center}

\begin{abstract}

We study the large~$N$ limit of the \emph{matrix valued} Gross-Neveu model in
$2<d<4$ dimensions. The method employed is a combination of the approximate
recursion formula of Polyakov and Wilson with the solution to the zero
dimensional large~$N$ counting problem of Makeenko and Zarembo. The model is
found to have a phase transition at a finite value for the critical temperature
and the critical exponents are approximated by $\nu = 1/(2(d-2))$ and
$\eta=d-2$. We test the validity of the approximation by applying it to the
usual vector models where it is found to yield exact results to leading order
in $1/N$.

\end{abstract}

\end{titlepage}

\section{Introduction}
\label{introduction}

In spite of more than twenty years of efforts, we are still lacking a
quantitative understanding of the large~$N$ limit of matrix valued field
theories in more than two dimensions. A solution to this problem would be
extremely welcome because of its twofold application to gauge theory on one
side (as pioneered by 't Hooft in~\cite{thooft1}), and string theory and random
surfaces on the other (see~\cite{review1} for a review of the solution to lower
dimensional systems and for a list of references).

This state of affairs is to be compared with our successful understanding of
vector valued models~\cite{review2}. The difference between these two problems
is more technical than conceptual; for vector models, where the number of
fields grows like $N$, we can use saddle point techniques, whereas this is not
directly possible if the number of fields grows like $N^2$. This difficulty
has been circumvented in dimension $d\leq 1$~\cite{brezin} and, for
gauge theory, $d=2$~\cite{thooft2} but the naive extension to
higher dimensions faces
formidable computational problems, although recently some progress has been
made
in this direction~\cite{master} by pursuing the master field
approach~\cite{witten}.

In a previous paper~\cite{myself}, we have proposed an approximation to the $d$
dimensional problem that combines the known solution to the corresponding zero
dimensional large~$N$ problem and the Polyakov-Wilson recursion
formula~\cite{wilson},~\cite{polyakov}. We have applied this approximation to
the three dimensional hermitian matrix model and found the values for the
critical exponents to be $\eta= 0.20$ and  $\nu= 0.67$.

In this paper we extend our investigation to the study of the \emph{matrix
valued} Gross-Neveu model in $2<d<4$ dimensions, described by the action
\beq
    S=\intx\tr\left(\pb\dirac\Psi +
    {\frac{u_0}{2N}}\pb\Psi\pb\Psi\right). \label{action}
\eeq
The action has the same form as the one for the original model~\cite{gross}
with the crucial difference that the fields are now arranged into a matrix
rather than a column vector. This means that the field content is, ignoring the
spin indices, $\pb_i^j(x), \Psi_k^l(x)$,  ($i,j,k,l = 1, \cdots, N$). It might
be
disturbing to some people to have fermionic fields living in fractional
dimensions but it is possible to make sense of them with the usual
prescriptions of first performing the calculations for the Dirac algebra and
then analytically continue in $d$. Each element of the matrix can be assumed to
be a two-component complex spinor but their Dirac algebra will never be written
out explicitly since, as we shall see, the only relevant properties are the
anticommuting nature of the fields and the existence of a discrete
$\mathbf{Z}_2$ symmetry preventing the occurrence of a bare mass term. This
symmetry can be viewed either as ``chirality'' near two dimensions or
``parity'' near three dimensions:
\beqs
      d=2&:&\qquad x\to x \quad \Psi\to \gamma_{\mathrm{ch}} \Psi
           \quad \pb \to -\pb \gamma_{\mathrm{ch}} \nonumber \\
      d=3&:&\qquad x\to -x \quad \Psi\to  \Psi \quad \pb \to -\pb.
      \label{chiralparity}
\eeqs

Our main results are the following. The fermionic matrix model has the same
qualitative phase structure as the ordinary vector model, with a phase
transition at some critical value for the inverse temperature
$u^*$ for $2< d <4$, with $u^*\to 0$ as $d\to 2$. The universal quantities at
criticality are however quite different, showing that the two models are not in
the same universality class. In particular, to leading order in $1/N$, the
``basic'' critical exponents for the matrix model within our approximation are
\beq
     \nu = {\frac{1}{2(d-2)}} \qquad \eta = d - 2, \label{nueta}
\eeq
(to be compared with $\nu = 1/(d-2)$ and  $\eta = 0$ for the vector model).

As we will argue towards the end of the paper, it might even be that these
exponents are closer to the actual value than one would generally expect from
an approximation of this kind. This is related to the existence of a small
parameter (the thickness of the integration shell in momentum space), under
which the recursion formula for the
fermionic model is better behaved than its
bosonic analogue.

The paper is organized as follows. In section~\ref{nature} we review the nature
of the approximation and show how it can be applied to various models. We shall
try to avoid too many repetitions; for more details see~\cite{myself}.
In section~\ref{appvector} we apply our approximation to the solvable vector
model and compare the results with the exact solution. We show that in a
particular limiting case it is possible to obtain exact results for the
exponents. A similar phenomenon has been found for the bosonic case
in~\cite{nishigaki} and we shall compare the two cases.
Section~\ref{appmatrix} is devoted to the study of the matrix model. The
recursion formula is obtained and the critical properties of the system are
investigated. In particular, the various critical exponents are computed for
$2 < d < 4$ within this approximation.
In section~\ref{conclusions} we conclude
with a discussion of our results and some future projects.
The relevant results for the corresponding zero dimensional vector and matrix
theories are summarized in appendix~\ref{appendixvect} and~\ref{appendixmatr}
respectively.

\section{The nature of the approximation.}
\label{nature}
The approximation we proposed in~\cite{myself} is a combination of the
Polyakov-Wilson approximate recursion formula~\cite{wilson}~\cite{polyakov} and
the solution to the appropriate zero dimensional large~$N$ counting problem (in
this particular case~\cite{fermizero}). To perform wilsonian renormalization on
a (euclidean) field theory one introduces an ultraviolet cut-off $\Lambda$,
that we shall always set to one by a choice of scale, and then integrates over
the Fourier components of the fields having momenta $\rho<p<1$ for some
$\rho\in]0,1[$. By rescaling the fields and the momenta so that the new action
has the same form as the original one, one obtains a recursion relation for the
coupling constants of the theory describing the critical behavior of the model.

In many cases the integration of the fast Fourier modes cannot be performed
exactly. The approximate recursion formula is perhaps the simplest, albeit
uncontrolled\footnote{We shall see later that perhaps, in some special cases,
there is a controlling small parameter.}, way to proceed beyond perturbation
theory.
It amounts to calculating all Feynman graphs by setting the incoming momenta
to zero and approximating the propagators and the loop integrals by two
constants, $b$ and $c$ respectively.  It is clear that this reduces the
integration problem to a zero dimensional counting problem, where the
dimensionality of space $d$ appears only parametrically through the rescaling
of fields and momenta at the last step of the renormalization\footnote{One
could also assume that the constant $c$ depends on the dimensionality of space
as in~\cite{myself}, but it turns out that the universal quantities are
independent of this choice, so we shall ignore this dependence altogether.}. An
important difference between the fermionic and the bosonic case is that in the
former we have to allow for $b$ to be complex; the actual phase of $b$ having
to be determined a posteriori by the requirement that the critical temperature
be real positive.

\begin{figure}

\begin{picture}(400,70)

\thicklines
\put(200,50){\circle{30}}
\put(185,50){\line(1,0){30}}
\thinlines
\put(150,50){\line(1,0){35}}
\put(215,50){\line(1,0){35}}

\end{picture}

\caption{\textsl{The setting sun is one of those
diagrams that are not properly
counted in the original recursion formula because it
has an odd number of external lines connecting at some vertex. This
distinction plays no role in our formulation.}}
\label{settingsun}
\end{figure}
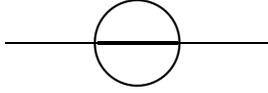

In its original formulation~\cite{wilson}~\cite{polyakov}, this approximation
was performed for a single component
bosonic field $\Phi(x)$. One wrote an integral
recursion relation for a function $v(\phi)$ of a real variable $\phi$ as the
approximation to the full effective potential $V[\Phi]$. This formulation has
the advantage of keeping track of the renormalization of all the ultralocal
terms of the kind $\intx \Phi(x)^n$,  but it has the drawback of ruling out
wave function renormalization from the onset (see however~\cite{golner}) and
excluding a large class of important diagrams such as the ``setting sun''
depicted in fig.~\ref{settingsun}

The power of combining the recursion formula with the large~$N$ limit is that
the integral for $v(\phi)$ can be performed explicitly ($\phi$ is now a vector
or matrix valued object) reducing the integral recursion relation to an
algebraic one that is \emph{ analytic} at the gaussian point because of the
nice convergence properties of large~$N$ integrals. All the leading diagrams in
$1/N$ are at least \emph{counted} properly, including the ``setting sun'' and
its friends. Wave function renormalization is easy to introduce in this scheme;
one simply isolates the contributing diagrams, i.e., those diagrams with two
external legs not joining at the same vertex, and approximates each of them as:
\beqs
        D(q^2) &=& D(0)+q^2 D^\prime (0)+\cdots\approx D(0)-
         q^2 b D(0)+\cdots
        \mbox{Êfor bosons}Ê\nonumber \\
        D(\qslash) &=& D(0) + \qslash D^\prime (0) +\cdots
        \approx D(0) - i \qslash \; b D(0) + \cdots
        \mbox{Êfor fermions}, \label{extraapprox}
\eeqs
where $b$ is the constant replacing the propagator as before. Replacing the
derivative by a constant is in the same spirit as the rest of the approximation
since the derivative ``adds'' a propagator to the integral of the Feynman
diagram. It is also the correct substitution on dimensional grounds.

In~\cite{myself} these approximations were applied to the hermitian matrix
model
\beq
      S[\Phi]=\frac{1}{2}\intx\tr\bigg((\nabla\Phi)^2 + r_0 \Phi^2\bigg) +
      \frac{u_0}{ N}\intx\tr \Phi^4,
\eeq
yielding the approximate recursion relations\footnote{In~\cite{myself} we had
set $b=1/(1+r)$ and $\rho=1/2$.}, ($l=0, 1, \cdots$)
\beqs
        r_{l+1}&=&4{\frac{r_l + 2(1+r_l)f_{r_0}(g)}{1+2f_{q^2}(g)}}\nonumber \\
        u_{l+1}&=&2^{4-d} u_l {\frac{1 + f_{u_0}(g)}{(1+2f_{q^2}(g))^2}},
        \label{boserg}
\eeqs
in terms of three known functions $f_{r_0}(g)$, $f_{u_0}(g)$ and $f_{q^2}(g)$
(mass, coupling constant and wave function renormalization respectively)
depending only on the coupling constant of the zero dimensional matrix
model~\cite{brezin}
\beq
        s(\phi) = \tr\left( \frac{1}{2}\phi^2 + \frac{g}{ N}\phi^4\right).
\eeq
The relation between $g$ and the physical quantities $r_l$ and $u_l$ was found
by analyzing the topology of the diagrams, (i.e., by counting the relative
number of loops, propagators and vertices), and it was found that
 \beq
         g(u_l, r_l) = \gur.
\eeq
By studying~(\ref{boserg}) it was possible
to estimate the critical exponents for
the three dimensional hermitian matrix model to be $\nu=0.67$ and $\eta=0.20$.

There are two main differences between the bosonic case of~\cite{myself} and
the fermionic one treated here, irrespective of whether they are vector or
matrix models. The first difference
is that the symmetry~(\ref{chiralparity}) prevents
a bare mass for the fermions\footnote{This symmetry is of course spontaneously
broken and the fermions acquire a mass, but all we need here is that no
Feynman graph gives rise to a term $\approx \pb\Psi$ after integration.}. The
second is that the four fermions interaction is \emph{ irrelevant}
at the gaussian point. The outcome of this is that the analogue of the pair of
equations~(\ref{boserg}) is a single recursion relation for $u_l$ and the
approximate renormalization group flow takes place on a line rather than on a
plane (see fig.~\ref{rgflow}) for which the gaussian point is attractive. For
both
vector and matrix models, we will show that there is another fixed point $u^*$
that is repulsive, hence corresponding to a phase transition, and compute the
critical exponents at that point. In a language perhaps more familiar to
particle physicists we could call the gaussian point infrared stable and $u^*$
ultraviolet stable. There is no possibility of confusion as long as we are
dealing with a one dimensional flow.

In~\cite{nishigaki} a similar calculation was performed for the bosonic vector
and matrix models where the thickness $1-\rho$ of the integration shell  was
kept arbitrary. It was found that the critical exponents were mildly dependent
on the unphysical value of $\rho$; an indication that some non universal
quantities had ``sneaked in'' during the approximation. By comparison with the
bosonic vector model, it was then argued that the exponents might become exact
in the limit $\rho\to 0$. We shall see that a similar phenomenon occurs here as
well, but for the opposite limit $\rho\to 1$. We try to give an explanation of
this curious fact in section~\ref{appmatrix} and~\ref{conclusions}.

\section{Application to the vector model.}
\label{appvector}
To test our technique and to gain some more insights, let us begin by
considering the original model~\cite{gross} with vector valued fermions $\pb_i,
\Psi^j$, ($i, j=1,\cdots,N$):
\beq
      S=\intx\left(\pb\dirac\Psi +
      {\frac{u_0}{2N}}(\pb\Psi)^2\right). \label{fermivector}
\eeq
This model is very well understood and its critical exponents have been
extensively studied~\cite{results}. To leading order they are $\eta=0$ and
$\nu=1/(d-2)$. The discrete chiral symmetry is always broken for $d=2$, where
the model is asymptotically free and the fermions are massive. For $2<d<4$, the
model has a phase transition at some finite value of the temperature, above
which chiral symmetry is restored.

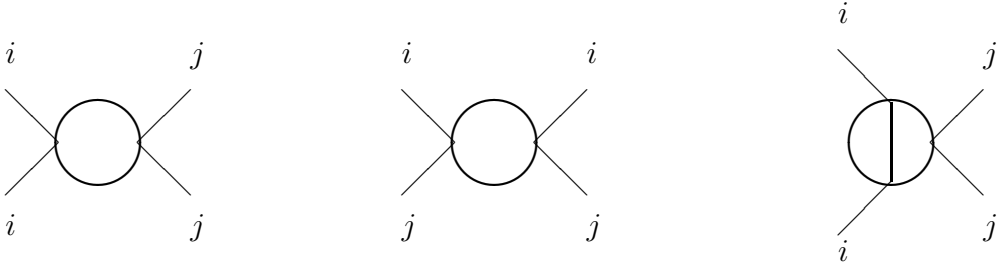
\begin{figure}

\begin{picture}(400,80)

\thicklines
\put(50,50){\circle{30}}
\put(200,50){\circle{30}}
\put(350,50){\circle{30}}
\put(350,35){\line(0,1){30}}
\thinlines
\put(35,50){\line(-1,1){20}}
\put(35,50){\line(-1,-1){20}}
\put(65,50){\line(1,1){20}}
\put(65,50){\line(1,-1){20}}
\put(185,50){\line(-1,1){20}}
\put(185,50){\line(-1,-1){20}}
\put(215,50){\line(1,1){20}}
\put(215,50){\line(1,-1){20}}
\put(350,65){\line(-1,1){20}}
\put(350,35){\line(-1,-1){20}}
\put(365,50){\line(1,1){20}}
\put(365,50){\line(1,-1){20}}
\put(15,80){$i$}
\put(15,15){$i$}
\put(85,80){$j$}
\put(85,15){$j$}
\put(165,80){$i$}
\put(165,15){$j$}
\put(235,80){$i$}
\put(235,15){$j$}
\put(330,95){$i$}
\put(330,5){$i$}
\put(385,80){$j$}
\put(385,15){$j$}

\end{picture}

\caption{\textsl{Among these three simple diagrams,
only the first one is leading.
The thin external lines represent the slow fields, the thick internal ones
the fast fields. The only vertex allowed between the two in a leading
diagram is $(\pb_s\Psi_s)(\pb_f\Psi_f)$. Of course, there can be any number
of vertices with four internal lines.}}
\label{cactuslike}
\end{figure}

Let us split the field $\Psi$ into the sum\footnote{To keep the notation
simple, we indicate only $\Psi$ and not $\pb$ but it should always be
understood that they are independent and subject to the same treatment.} of
$\Psi = \Psi_s + \Psi_f$ carrying only slow ($0<p<\rho$) and fast ($\rho<p<1$)
Fourier components and define, as usual,
\beq
       S[\Psi_s+\Psi_f]=S[\Psi_s] + \sigma[\Psi_s, \Psi_f] + S[\Psi_f].
       \label{splitting}
\eeq
The term $\sigma[\Psi_s, \Psi_f]$ is the part of $S$ that does not factorize
under the decomposition, namely,
\beq
       \sigma[\Psi_s, \Psi_f]  = {\frac{u_0}{N}}\intx
       (\pb_s\Psi_s)(\pb_f\Psi_f) +\cdots,
\eeq
where the dots represent terms like,
for instance, $(\pb_s\Psi_f)(\pb_s\Psi_f)$ that
would give subleading contribution in $1/N$ after integrating out the fast
modes. This fact should be very familiar to the large~$N$ experts given the
index structure of the ``cactus'' diagrams
(see fig.~\ref{cactuslike}).

We then define a new action $S^\prime$ obtained by integrating out the fast
modes\footnote{The use of the measure $\exp (S)$ instead of $\exp (-S)$ is
allowed for Grassmann integrals.}:
\beq
       e^{S^\prime[\Psi_s]} = \int  \mathcal{D}\Psi_f \;e^{S[\Psi_s+\Psi_f]}=
       e^{S[\Psi_s]} \int \mathcal{D}\Psi_f\;
            e^{\sigma[\Psi_s, \Psi_f] + S[\Psi_f],},
\eeq
and perform the integral to leading order in $1/N$, keeping the terms
$O(\Psi_s^4)$ to \emph{all orders} in perturbation theory, to obtain
\beqs
       S^\prime[\Psi_s] &=&S[\Psi_s] + {\frac{u_0^2}{2N}}\intxy
       (\pb_s\Psi_s)(x)
       (\pb_s\Psi_s)(y)\times\nonumber \\
       &&\bigg( <\frac{1}{N}(\pb_f\Psi_f)(x)
       (\pb_f\Psi_f)(y)>_{\mathrm{conn.}}-
       <\frac{1}{N}\pb_f(x)\Psi_f(y)>^2_{\mathrm{conn.}} \bigg),
       \label{exactprime}
\eeqs
where the connected Green functions are defined with the full measure $\exp
(S[\Psi_f])$ to all orders in $u_0$. Eq.~(\ref{exactprime}) is the exact
correction to the four fermions term to leading order in $1/N$. We claim that
if we approximate~(\ref{exactprime}) by making the assumption listed
in section~\ref{nature} we obtain:
\beq
       S^\prime[\Psi_s] = \intx\left(\pb_s\dirac\Psi_s +
      {\frac{u_0}{2N}} {\frac{\Gamma_4(g_0)}{g_0}}(\pb_s\Psi_s)^2\right),
      \label{sprimevector}
\eeq
where $\Gamma_4$ is the amputated one particle irreducible four point function
for the analogous zero dimensional vector model $s(\psi)=\pbs\psi + g/(2N)
(\pbs\psi)^2$ (see appendix~\ref{appendixvect}), and the physical coupling
constant $u_0$ is
related to the zero dimensional one by
\beq
    g_0 = u_0 c b^2.
\eeq

A very explicit way to go from~(\ref{exactprime}) to~(\ref{sprimevector}) is to
make the approximation directly in~(\ref{exactprime}) and then use the zero
dimensional Schwinger-Dyson equation
\beq
    \Gamma_4(g) = g + g^2(C_4(g) - C_2^2(g)).
\eeq

One can check directly the validity of~(\ref{sprimevector})
expression by expanding $S^\prime$
to the first few orders in perturbation theory. It should be clear however
that, after setting the external momenta to zero, the only surviving diagrams
are the one particle irreducible ones\footnote{All internal propagators vanish
for $q<\rho$.} and that the relative powers $L$, $P$ and $V$ of loops,
propagators and vertices are controlled by the following topological formulas
valid for the four point function:
\beq
        2V= 2 +P \quad\mbox{and}\quad V-P+L=1. \label{topofour}
\eeq
Solving for $P$ and $L$ we see that a diagram with $V$ vertices gives a
contribution proportional to
\beq
        u_0^V \times b^{2(V-1)} \times c^{V-1} = u_0 g_0^{V-1}
\eeq
as properly encoded in the power series of~(\ref{sprimevector}).

The final step is to rescale coordinates and fields so that
action~(\ref{sprimevector}) has the same cut-off in dimensionless units. Since
no wave function renormalization appears to leading order for vector models,
the rescaling is the ``classical'' one: $x\to \rho^{-1} x$ and $\Psi_s\to
\rho^{(d-1)/2} \Psi$ to get the new coupling constant
\beq
        u_1 = \rho^{d-2} u_0 \times {\frac{\Gamma_4(g_0)}{g_0}}.
\eeq
It is actually more convenient to multiply both sides by $c b^2$ and express
everything in terms of $g_l = u_l c b^2$, ($l= 0, 1, 2, \cdots$). Thus, the
expression for the approximate recursion formula for the vector valued
Gross-Neveu model becomes
\beq
        g_{l+1} = \rho^{d-2} \Gamma_4(g_l). \label{vectorgrflow}
\eeq

\begin{figure}
\epsfig{file=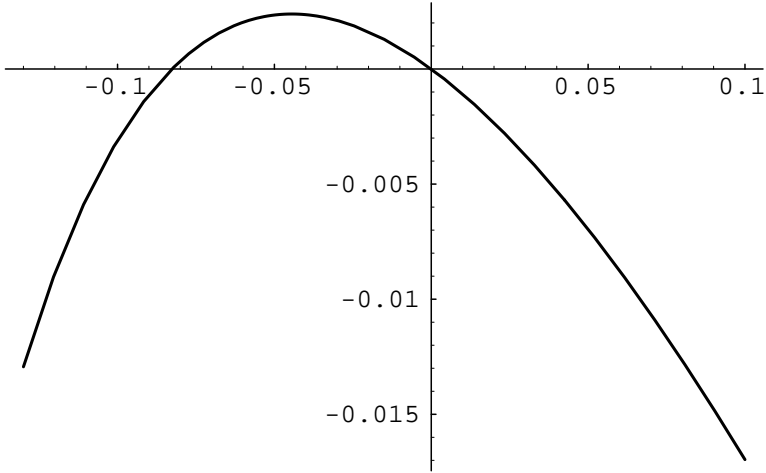}
\caption{\textsl{Plot of eq.~(\ref{vectorgrflow}). For clarity, we plot the
\emph{difference}
$\rho^{d-2}\Gamma_4(g) - g$ vs. $g$ for $d=3$ and
$\rho=0.9$. The function $\Gamma_4$ is real for all
$g > -1/4$; the range shown is $-0.13<g<+0.1$.}}
\label{plotvect}
\end{figure}

\begin{figure}

\begin{picture}(400,70)(30,0)

\thicklines
\put(100,30){\line(1,0){300}}
\put(200,30){\circle*{5}}
\put(300,30){\circle*{5}}
\put(150,30){\line(-1,1){10}}
\put(150,30){\line(-1,-1){10}}
\put(250,30){\line(1,1){10}}
\put(250,30){\line(1,-1){10}}
\put(350,30){\line(-1,1){10}}
\put(350,30){\line(-1,-1){10}}
\put(185,50){$u=0$}
\put(285,50){$u=u^*$}

\end{picture}

\caption{\textsl{The qualitative behavior of the renormalization group flow
for both vector and matrix valued fermionic models. The arrows represent the
change in the physical coupling $u_l$ under iteration $l\to l+1$.}}
\label{rgflow}
\end{figure}
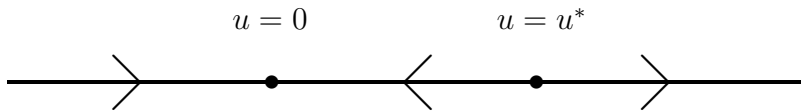

The behavior of~(\ref{vectorgrflow}) is shown in fig.~\ref{plotvect}. There are
two fixed points; the gaussian one and $g_l=g_{l+1}=g^*$. What is important is
that $g^*$ is unstable under $g_l \to g_{l+1}$ thus defining a phase
transition. This is consistent with the fact that the four fermions
term is irrelevant at the origin and makes the renormalization group flow look
as shown in fig.~\ref{rgflow} for the \emph{physical} coupling constant $u$.

The fact that the critical value of $g$ is negative
forces us to choose $b$ to be
purely imaginary in order to have a positive critical inverse temperature
$u^*$. There is no danger in doing this as $b$ does not have any direct
physical meaning. One can solve~(\ref{vectorgrflow}) explicitly for the
critical point $g^*$ and compute the exponent $\nu$ at that point by
linearization~\cite{wilson}:
\beq
       \nu = - {\frac{\log \rho}{\log (\rho^{d-2} \Gamma_4^\prime(g^*) )}} =
       - {\frac{\log \rho}{\log(2 - 3 \rho^{2-d} + 2 \rho^{4-2d})}}.
\label{nuvectrho}
\eeq
The exponent $\nu$ is always positive and independent on the unphysical
quantities $b$ and $c$ but still depends on the (also unphysical) parameter
$\rho$. We must admit that for, say, $\rho=1/2$,
formula~(\ref{nuvectrho}) does
not do a very good job, giving, for instance,
$\nu = 1/2$ at $d=3$. However, as we let the integration shell become thinner,
i.e., $\rho\to 1$, formula~(\ref{nuvectrho}) becomes exact, i.e., $\nu\to
1/(d-2)$. This is the analogue phenomenon found in~\cite{nishigaki} for the
bosonic case. Its occurrence for vector models is not too surprising. For
vector models, the leading value for $\nu$ is obtained by a one loop
calculation, the ``gap equation''. If there is a limiting value for $\rho$ for
which the critical coupling $u^*$ becomes small, the two calculations are bound
to coincide. If anything, this observation provides a useful check of our
calculation. Also, the fact that for bosons~\cite{nishigaki} the exact value is
reached as $\rho\to 0$ rather that $\rho\to 1$ is a direct expression of the
fact that, there, $(\Phi^2)^2$ is relevant.

In the fermionic case, we can think
of $\rho$ as a sort of ``control'' parameter\footnote{To be precise,
the control parameter is the thickness of the integration
shell $1-\rho$.} for the (generically uncontrolled)
approximate recursion formula, in the sense that by choosing a very thin
integration shell in momentum space, we can keep the critical point near the
origin and get exact results for the exponents. Of course, for a thin shell the
critical point is kept near the origin in the bosonic case as well, but, there,
the exact exponents are recovered in the opposite limit because of the
different scaling dimension of the interaction; the interpretation of
$\rho$ as a control parameter becomes then more problematic.
We now move to the case of
matrix valued fields, where we shall see that the difference between the
bosonic and fermionic case is even more significant.

\section{The matrix valued Gross-Neveu model.}
\label{appmatrix}

The model of real interest in this paper is described by the
action~(\ref{action}) presented in the introduction. Since the solution to the
corresponding zero dimensional problem is known~\cite{fermizero}, there are no
obstacles to extending our technique to this system. We can still decompose
$\Psi$ and $\pb$ into fast and slow modes,
as before, with the assumption that
all fields are matrix valued. We can still write equation~(\ref{splitting}),
but now
\beqs
        \sigma[\Psi_s, \Psi_f] &=& {\frac{u_0 }{2 N}} \intx \tr \bigg(
        2\pb_s\Psi_f\pb_f\Psi_f + 2\pb_f\Psi_s\pb_f\Psi_f\nonumber \\&& +
        2\pb_s\Psi_s\pb_f\Psi_f + 2\pb_s\Psi_f\pb_f\Psi_s +
         \pb_s\Psi_f\pb_s\Psi_f\nonumber \\&& + \pb_f\Psi_s\pb_f\Psi_s +
        2\pb_f\Psi_s\pb_s\Psi_s + 2\pb_s\Psi_f\pb_s\Psi_s \bigg).
        \label{sigmamatr}
\eeqs

Although we will never need this in practice, it should be mentioned that
some of the terms in~(\ref{sigmamatr}) do not contribute to leading order: the
fifth and sixth terms give rise to diagrams that are subleading in $1/N$ and
the last two terms do not contribute to the four point function because they
only appear in one particle \emph{reducible} diagrams that vanish by momentum
conservation. Hence, only the first four terms in~(\ref{sigmamatr}) survive.
One could perform the integral over $\Psi_f$ and find the form of $S^\prime$ in
terms of the exact connected Green functions. We skip this tedious calculation
because it is a repetition of what
was done in section~\ref{appvector} and because
a detailed calculation for the bosonic case is presented in~\cite{myself}.
Instead, we shall argue directly from the diagrams what the answer should be.

We need to compute coupling constant \emph{and} wave function
renormalization\footnote{Contrary to the vector model, wave function
renormalization appears to leading order in the matrix model.}. Coupling
constant renormalization is essentially the same story as for the vector model;
by setting the external momenta to zero, we select the one particle irreducible
diagrams, i.e., $u_0 \to u_0 \Gamma_4(g_0)/g_0$ where now $\Gamma_4$ is the
vertex function of the fermionic matrix model, computed in
appendix~\ref{appendixmatr}. The relation $g_0=u_0 c b^2$ is unchanged because
the topological formulas~(\ref{topofour}) are model
independent\footnote{In~(\ref{topofour}), the integer $L$ refers to the number
of momentum loops, not color loops.}.

\begin{figure}

\begin{picture}(400,200)

\thicklines
\qbezier(100,40)(30,70)(100,100)
\qbezier(100,40)(170,70)(100,100)
\put(100,97){\circle*{15}}
\thinlines
\put(100,40){\line(1,-1){20}}
\put(100,40){\line(-1,-1){20}}
\thicklines
\qbezier(250,70)(250,120)(300,120)
\qbezier(250,70)(250,20)(300,20)
\qbezier(300,120)(350,120)(350,78)
\qbezier(300,20)(350,20)(350,62)
\put(300,120){\circle*{15}}
\put(300,70){\circle*{15}}
\put(300,20){\circle*{15}}
\put(250,70){\line(1,0){92}}
\put(350,70){\circle{15}}
\thinlines
\put(250,70){\line(-1,0){30}}
\put(358,70){\line(1,0){30}}
\put(30,150){\textbf{A}}
\put(250,150){\textbf{B}}

\end{picture}

\caption{\textsl{The full contribution to
wave function renormalization comes from
diagram B. Diagrams like A should be ruled out by symmetry considerations
but they already vanish in the zero dimensional matrix theory. In this figure,
dark circles represent the connected Green functions and
the light circle the irreducible four point vertex.}}
\label{wavefunction}
\end{figure}
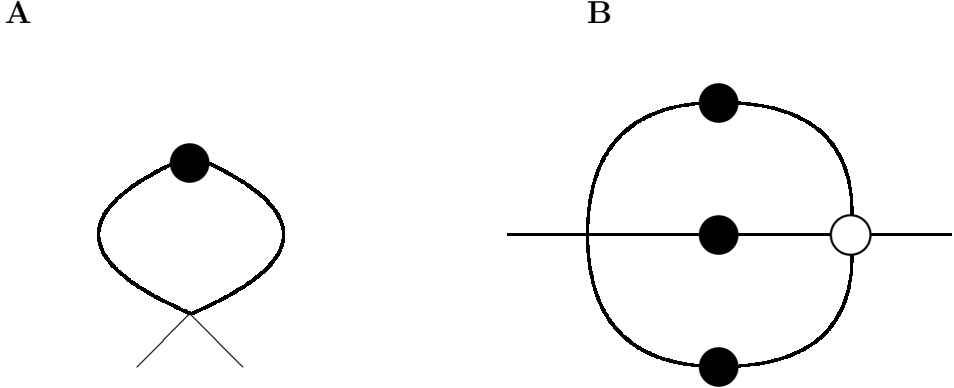

As for wave function renormalization, we need to isolate the contributing
diagrams and apply the approximation~(\ref{extraapprox}). We let $q\approx 0$
and analyze those diagrams that retain some dependence on $q$ in this limit.
Those are all the one particle irreducible diagrams with two external legs
\emph{not} connected at the same vertex, as shown in fig.~\ref{wavefunction}B.
All the diagrams that are not one particle irreducible vanish by momentum
conservation because the internal propagators are non zero only for
$\rho< q<1$,
and those diagrams where the two external lines connect at the same vertex
(fig.~\ref{wavefunction}A) do not have any dependence on $q$. Recalling the
symmetry~(\ref{chiralparity}), we expect these diagrams to vanish anyway in the
exact solution, together with the $q$ independent part of all other two point
functions.

When using the zero dimensional results, the symmetry~(\ref{chiralparity}) must
be imposed ``by hand''\footnote{There is a zero dimensional $\mathbf{Z}_4$
discrete symmetry $\Psi\to \pb$, $\pb\to - \Psi$, but this is not what we want
because, for instance,  the term $\tr (\pb\Psi\pb\Psi)$ is not invariant under
this symmetry.}. However, for the case of the matrix model we encounter a
surprise: diagram~\ref{wavefunction}A vanishes already in zero
dimensions because of the color algebra. This fact can be easily checked to one
loop in perturbation theory:
\beq
     <\pb_i^j \Psi_k^l \tr \pb\Psi\pb\Psi >|_0 =N \delta_i^l\delta^j_ k
     (1 + 1 - 1 - 1) = 0,
\eeq
and it is related to another curious property of the fermionic matrix model
described in appendix~\ref{appendixmatr}. As a result, the Schwinger-Dyson
equations allow us to write the contribution of diagram~\ref{wavefunction}B
in the much simpler form:
\beq
       1 - g C_2^3(g) \Gamma_4(g)  = 2 - \Gamma_2(g),
\eeq
yielding the integral over $\Psi_f$
\beq
     S^\prime[\Psi_s] = \intx\tr\left( (2 - \Gamma_2(g_0))
     \pb_s\dirac\Psi_s +
     {\frac{u_0}{2N}} {\frac{\Gamma_4(g_0)}{g_0}}
     \pb_s\Psi_s\pb_s\Psi_s\right). \label{sprimematrix}
\eeq

The relative powers of $u_0$, $b$ and $c$ can be checked as before; the
topological formulas that apply to the two point function are
\beq
        2V= 1 + P \quad\mbox{and}\quad V-P+L=1, \label{topotwo}
\eeq
yielding a contribution for a diagram with $V$ vertices proportional to $u_0^V
c^V b^{2V -1}$. Multiplying by the extra propagator $b$, to approximate the
derivative as in~(\ref{extraapprox}), gives a contribution to wave function
renormalization proportional to $u_0^V c^V b^{2V} = g_0^V$.

The renormalization group recursion relations are obtained as before by
rescaling
$x\to\rho^{-1}x$ and $\Psi_s \to \rho^{(d-1)/2}(2-\Gamma_2(g_0))^{-1/2}\Psi$
and redefining the new coupling
\beq
        u_1 = \rho^{d-2} u_0 \times
       {\frac{\Gamma_4(g_0)}{g_0(2 - \Gamma_2(g_0))}}. \label{nqf}
\eeq
Better yet, we multiply~(\ref{nqf}) by $c b^2$ to obtain, for $l=0,1\cdots$,
\beq
        g_{l+1} = \rho^{d-2}{\frac{\Gamma_4(g_l)}{2 - \Gamma_2(g_l)}} \equiv
        \rho^{d-2}R(g_l), \label{matrixrg}
\eeq
where the last equation defines $R(g)$.

The form of $R(g)$ can be found after a fair amount of tedious algebra to be
\beq
       R(g) = \frac{g\beta(\beta+3)(\beta +6)^2}
        {(\beta + 4)^3(2\beta^2 + 9\beta +6)},
\eeq
where $\beta(g)$ is the solution to $g^2\beta^4 - \beta^2 -  8\beta  - 12=0$,
$\beta(0)=-2$ necessary to satisfy the one cut ansatz to the Riemann-Hilbert
problem as described in appendix~\ref{appendixmatr}.

\begin{figure}
\epsfig{file=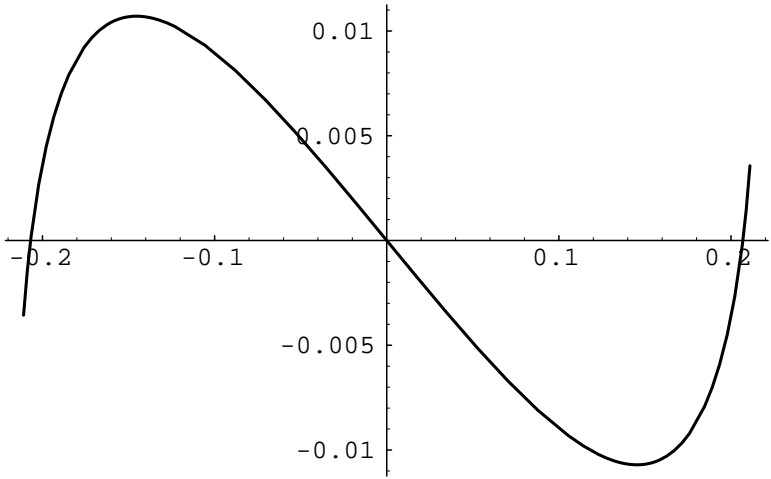}
\caption{\textsl{Plot of eq.~(\ref{tildergn}). For clarity,
we plot the \emph{difference}
$\rho^{d-2}\tilde R(\tilde g) - \tilde g$ vs. $\tilde g$ for $d=3$ and
$\rho=0.9$. The function $\tilde R$ is real for $|\tilde g|<0.211$.}}
\label{plotmatr}
\end{figure}

By setting $g_l=g_{l+1}=g^*$ we see that~(\ref{matrixrg}) admits two purely
imaginary solutions $g^*=\pm i \tilde g^*$. This is fairly easy to show: the
function $R(g)$ is odd ($\beta$ is even) and real near the origin, where it has
a power series $R(g) = g - g^3 + O(g^5)$. Hence, for $\rho\to 1^-$ we obtain
$g^{*2}\approx 1 - \rho^{2-d} < 0$ leading to two imaginary solutions. This can
be seen more generally by setting $R(i\tilde g) = i\tilde R(\tilde g)$. By
virtue of $R(g)$ being real and odd, $\tilde R(\tilde g)$ is also real and odd
and eq.~(\ref{matrixrg}) becomes
\beq
     \tilde g_{l+1} = \rho^{d-2}\tilde R(\tilde g_l).
     \label{tildergn}
\eeq
The plot of~(\ref{tildergn}) is shown in fig.~\ref{plotmatr} for the region in
which $\tilde R$ is real. One can see that there are only two $\mathbf{Z}_2$
symmetric non trivial solution, both of which
are unstable. The requirement that
$u^*$ be real positive forces us to fix $b^2$ to be purely imaginary and to
discard one of the two solution as unphysical. The renormalization group flow
involving the gaussian point and $u^*$ is qualitatively the same as the one
depicted in fig.~\ref{rgflow}.

There are other solutions to~(\ref{matrixrg}) in the whole complex plane but
the above solution
is the only one that can be really trusted by virtue of reducing to the
gaussian when the integration shell becomes thinner ($\rho\to 1$). In fact we
see that, just as for the fermionic vector model in section~\ref{appvector},
the parameter $\rho$ provides a way of controlling the approximation by keeping
the critical point near the gaussian one for thin integration shells. The
situation is more favorable than in the bosonic case~\cite{nishigaki} because
it is in this region that the exponents for the fermionic vector model become
exact.
It is possible to linearize~(\ref{matrixrg}) numerically and extract the
exponent $\nu$. However, as we have argued that the results should be reliable
for $\rho\approx 1$, it is enough to consider the first non trivial term in the
Taylor expansion near the origin:
\beq
     g_{l+1} \approx \rho^{d-2}(g_l - g_l^3).
\eeq
To illustrate why we expect a universal behavior different
from the one of the vector
model, let us recall that in the vector case the Tailor expansion
of~(\ref{vectorgrflow}) contains a quadratic term:
$g_{l+1} \approx  \rho^{d-2}(g_l - g_l^2)$. In the most general case
\beq
    g_{l+1} \approx \rho^{d-2}(g_l + A g_l^n),
\eeq
one obtains
\beq
     \nu \approx -\frac{\log\rho}{\log(n - (n-1)\rho^{d-2})} \to
     \frac{1}{(n-1)(d-2)} \quad\mbox{as}\quad \rho\to 1.
\eeq
For $n=2$ we recover the exponent for the vector model and for $n=3$ we prove
our first claim on the matrix model: $\nu = 1/(2(d-2))$.

In a similar way, it is possible to evaluate the wave function renormalization
at the critical point and to extract the exponent
\beq
     \eta = -\frac{\log (2 - \Gamma_2(g^*))}{\log\rho} \to d-2
     \quad\mbox{as}\quad \rho\to 1.
\eeq
If we are willing to be optimistic, we can use the combined values of $\nu$ and
$\eta$ to compute the other magnetic exponents from the well known scaling
relations:
\beq
     \alpha=\frac{3d-8}{2(d-2)},\quad\beta=\frac{1}{2},
     \quad\gamma=\frac{4-d}{2(d-2)}\quad\mbox{and}\quad\delta=\frac{2}{d-2}.
\eeq
It should be possible to test the validity of some of these results by a
computer simulation.

\section{Conclusions.}
\label{conclusions}

We have shown how to compute the critical exponents for the matrix valued
Gross-Neveu model by using a combination of the approximate recursion formula
and the zero dimensional large~$N$ limit. The success of this technique in
predicting the exponents for the ordinary vector models (both bosonic and
fermionic) gives us more confidence in the approximation. The recursion
relation for the fermionic models is even better
behaved that for the bosonic ones; the
thickness of the integration shell controls the distance of the critical point
from zero.

Since we have used a Taylor series expansion to obtain the values of the
critical exponents, one might be tempted to say that a
perturbative calculation should suffice.
This is not the case because the existence of a non
trivial fixed point can never be established by perturbation theory alone; one
must rely on some non perturbative treatment such as the
one used in this paper.
However, it is true that for a certain range of the parameter $\rho$, the fixed
point falls near the origin and then there is no harm in doing a Taylor
expansion. In a sense, the situation is similar that in the
$\epsilon$~expansion~\cite{wf} where the controlling parameter $\epsilon$ can
be chosen in such a way that the fixed point also falls near the origin, thus
allowing one to expand in powers of the coupling constant. The parameter $\rho$
plays a role similar to $\epsilon$ but only for fermionic models. One crucial
difference is that $\epsilon$ is a physical parameter (essentially, the
dimension of space) whereas $\rho$ is not.
It would be interesting to  develop a self consistent way to
show that is is possible to
neglect higher powers of $\Psi$ from the exact renormalization group equations.

Two directions worth exploring in the future are the connection with randomly
triangulated surfaces and the application of similar methods to gauge theories.
In both cases, the simple approximations used in this paper need to be modified
for many reasons; the critical points studied here and in~\cite{myself} are
not the ones corresponding to the continuum limit of the string partition
function and gauge symmetry is not preserved in the presence of a sharp
cut-off. A solution to either of these problems, allowing one to carry out the
spirit of the approximation for these systems, would be extremely welcome.

\section*{Acknowledgements.}

We would like to thank J. Ambj{\o}rn, P. Damgaard, S. Nishigaki and
G. Semenoff for discussions.

\appendix

\section{The zero dimensional fermionic vector model.}
\label{appendixvect}

In this appendix we quickly review the relevant results for the fermionic
vector model. We will be brief but try to be self contained; for more details
see~\cite{fermizero}. The great simplification for the vector models is that
their Schwinger-Dyson equations can be solved exactly in a closed form,
contrary to the matrix model case where, even after imposing the
Schwinger-Dyson equations, one is still left with some undetermined
correlators.
Hence, for the vector model, we can obtain the
quantities we need without doing any real work!

For any quantity $\mathcal{O}$ we define
\beq
     <\mathcal{O}>=\frac{\int d\psi\; d\pbs\; \mathcal{O}e^{\pbs\psi +
                \frac{g}{2N}(\pbs\psi)^2} }
     {\int d\psi\; d\pbs\;e^{\pbs\psi +
                \frac{g}{2N}(\pbs\psi)^2}}.
\eeq
It is a simple matter to check that the first two Schwinger-Dyson equations
read
\beqs
      <\frac{1}{N}\pbs\psi> + g <\frac{1}{N^2}(\pbs\psi)^2> -1 &=&0\nonumber\\
      <\frac{1}{N^2}(\pbs\psi)^2> + g <\frac{1}{N^3}(\pbs\psi)^3> -
      \left(1 - \frac{1}{N}\right)<\frac{1}{N}\pbs\psi> &=& 0. \label{sdvect}
\eeqs
Eq.~(\ref{sdvect}) is exact, i.e., valid for all finite $N$'s.
Now we define the connected correlation functions by cluster decomposition:
\beqs
     <\pbs^i\psi_j> &=& <\pbs^i\psi_j>_{\mathrm{conn.}} \nonumber\\
     <\pbs^i\psi_j\pbs^k\psi_l> &=& <\pbs^i\psi_j>_{\mathrm{conn.}}
     <\pbs^k\psi_l>_{\mathrm{conn.}} - \mbox{ cross term } \nonumber\\
     &&+<\pbs^i\psi_j\pbs^k\psi_l>_{\mathrm{conn.}} \nonumber\\
     <\pbs^i\psi_j\pbs^k\psi_l\pbs^m\psi_n> &=& <\pbs^i\psi_j>_{\mathrm{conn.}}
     <\pbs^k\psi_l>_{\mathrm{conn.}}<\pbs^m\psi_n>_{\mathrm{conn.}} \pm
     \mbox{ five c.t.}\nonumber\\
     &&+ <\pbs^i\psi_j>_{\mathrm{conn.}}
     <\pbs^k\psi_l\pbs^m\psi_n>_{\mathrm{conn.}}\pm\mbox{ eight c.t. }
     \nonumber\\ &&+<\pbs^i\psi_j\pbs^k\psi_l\pbs^m\psi_n>_{\mathrm{conn.}}.
     \label{clustervect}
\eeqs
For our purposes, we need to keep track of terms of order $1/N$
in~(\ref{clustervect}) and this can be done by introducing three quantities
$C_2(g)$, $\tilde C_2(g)$ and $C_4(g)$ through
\beqs
     <\pbs^i\psi_j>_{\mathrm{conn.}}&=&\delta^i_j\left(C_2(g) +
     \frac{1}{N} \tilde C_2(g)\right) + O(\frac{1}{N^2}) \nonumber \\
     <\pbs^i\psi_j\pbs^k\psi_l>_{\mathrm{conn.}} &=& \left(
     \delta^i_j\delta^k_l-\delta^i_l\delta^k_j\right)
     \frac{1}{N} C_4(g) + O(\frac{1}{N^2}) \nonumber \\
     <\pbs^i\psi_j\pbs^k\psi_l\pbs^m\psi_n>_{\mathrm{conn.}} &=&
     O(\frac{1}{N^2}). \label{expandvect}
\eeqs
Summing over the color indices in~(\ref{clustervect}) and substituting
into~(\ref{sdvect}) we obtain, after keeping careful track of all powers of
$1/N$,
\beqs
     &&C_2 + g C_2^2 - 1 = 0 \nonumber \\
     &&\tilde C_2 + g(2 C_2 \tilde C_2 + C_4 - C_2^2) = 0 \nonumber \\
     &&2 C_2 \tilde C_2 + C_4 - C_2^2 + 3gC_2(C_2 \tilde C_2 + C_4 - C_2^2)
     - \tilde C_2 + C_2 = 0. \label{systemvect}
\eeqs
Eq.~(\ref{systemvect}) forms a closed system of equations that can be solved
for the quantities that we need:
\beqs
     C_2(g) &=& \frac{\sqrt{1 + 4g} - 1}{2g} \nonumber \\
     C_4(g) &=& \frac{(1+2g -\sqrt{1+4g})^2}{2 g^2 (1+4g-\sqrt{1+4g})}.
\eeqs
By a decomposition similar to~(\ref{clustervect}) it can be immediately checked
that the relations between the connected Green functions $C_2$ and $C_4$ and
the vertex functions $\Gamma_2$ and $\Gamma_4$ are the usual ones:
\beqs
     \Gamma_2(g) &=& C_2(g)^{-1} = \frac{1 + \sqrt{1 + 4g}}{2}\nonumber \\
     \Gamma_4(g) &=& C_4(g) C_2(g)^{-4} =\frac{g}{2}\left(1 +
          \frac{1}{\sqrt{1 + 4g}} \right),
\eeqs
where the only unusual thing is, perhaps, the relative sign between $C_4$ and
$\Gamma_4$ that is dictated by the choice of the measure $\exp(+S)$. The vertex
function $\Gamma_4$ is the one used in the computation for the vector model in
section~\ref{appvector}.

\section{The zero dimensional fermionic matrix model.}
\label{appendixmatr}

We now move on to the fermionic matrix model for which we have to do a bit more
work since the Schwinger-Dyson equations do not close.  For any quantity
$\mathcal{O}$ we now define
\beq
     <\mathcal{O}>=\frac{\int d\psi\; d\pbs\; \mathcal{O}e^{\tr\left(\pbs\psi +
                \frac{g}{2N}\pbs\psi\pbs\psi\right) } }
                {\int d\psi\; d\pbs\; e^{\tr\left(\pbs\psi +
                \frac{g}{2N}\pbs\psi\pbs\psi\right) } } .
\eeq
The only Schwinger-Dyson equation we shall need is
\beq
    <\frac{1}{N^2}\tr(\pbs\psi)> + g <\frac{1}{N^3}\tr(\pbs\psi\pbs\psi)> -1=0.
    \label{sdmatr}
\eeq
The trick of cluster decomposition of appendix~\ref{appendixvect} will not work
here because the connected components are of the same order as the disconnected
ones. Instead, we follow~\cite{fermizero} and define a density\footnote{A
matrix with anticommuting elements cannot be diagonalized and $\rho$ no longer
has the meaning of eigenvalue density.} $\rho(\mu)$
\beq
     <\frac{1}{N^{k+1}}\tr\left((\pbs\psi)^k\right)> =
     \int d\mu\;\rho(\mu) \mu^k,
\eeq
and a generating function $\omega(z)$
\beq
     \omega(z)=<\frac{1}{N}\tr\left(\frac{1}{z - \frac{\pbs\psi}{N}}\right)> =
     \int d\mu\;\frac{\rho(\mu)}{z-\mu}.
\eeq
We shall not repeat the argument here but it is well known~\cite{fermizero}
that the density $\rho(\mu)$ is the same as the density for the \emph{Penner
model}~\cite{penner}, i.e., it has support on a finite complex
arch on which it satisfies
\beq
     \frac{1}{\lambda}-\frac{1}{2} -\frac{g\lambda}{2} =
     \wp\int d\mu\;\frac{\rho(\mu)}{\lambda-\mu}. \label{riemannhilbert}
\eeq
The one cut solution to the Riemann-Hilbert problem~(\ref{riemannhilbert}) is:
\beq
    \omega(z) = \frac{1}{z} - \frac{1}{2} - \frac{gz}{2} +
    \left(\frac{g}{2} - \frac{1}{\beta z}\right)\sqrt{z^2 +
    \alpha z + \beta^2},
\eeq
where $\alpha$ and $\beta$ are determined by the asymptotic condition
$\omega(z)\to 1/z$ at infinity. The quantity $\beta$ is the same as the one
used in section~\ref{appmatrix}.
After some tedious algebra, it is possible to eliminate $\alpha$ from the
problem and to write the two point function as
\beq
      <\frac{1}{N^2}\tr(\pbs\psi)> = -\frac{\beta(\beta+4)}{\beta+6},
\eeq
where $\beta$ is the solution to $g^2\beta^4 - \beta^2 - 8\beta -12 =0$ with
the initial condition $\beta= -2$~at~$g=0$ necessary to ensure the proper
gaussian limit. The four point function can be read off from~(\ref{sdmatr}).

To conclude, we wish to show how all this is related to the connected Green
functions $C_2$ and $C_4$ and the vertex functions $\Gamma_2$ and $\Gamma_4$.
This is achieved by simple cluster decomposition:
\beqs
     <\pbs^i_j\psi^k_l> &=& <\pbs^i_j\psi^k_l>_{\mathrm{conn.}} \nonumber\\
     <\pbs^i_j\psi^k_l\pbs^m_n\psi^p_q> &=& <\pbs^i_j\psi^k_l>_{\mathrm{conn.}}
     <\pbs^m_n\psi^p_q>_{\mathrm{conn.}} \nonumber\\
      && -<\pbs^i_j\psi^p_q>_{\mathrm{conn.}}
      <\pbs^m_n\psi^k_l>_{\mathrm{conn.}} \nonumber\\
      &&+<\pbs^i_j\psi^k_l\pbs^m_n\psi^p_q>_{\mathrm{conn.}}
      \label{clustermatr}
\eeqs
and by the usual definition
\beqs
     <\pbs^i_j\psi^k_l>_{\mathrm{conn.}}&=&\delta^i_l\delta^k_j C_2(g)
     + O(\frac{1}{N}) \nonumber \\
     <\pbs^i_j\psi^k_l\pbs^m_n\psi^p_q>_{\mathrm{conn.}} &=& \left(
     \delta^i_l\delta^k_j\delta^m_q\delta^p_n -
     \delta^i_q\delta^p_j\delta^m_l\delta^k_n\right)
     \frac{1}{N} C_4(g) + O(\frac{1}{N^2}).
     \label{expandmatr}
\eeqs
Substituting~(\ref{expandmatr}) into~(\ref{clustermatr}) and taking the trace,
we see that the disconnected piece
drops out because of the antisymmetry in the
color indices yielding
\beqs
       C_2(g)&\equiv&<\frac{1}{N^2}\tr(\pbs\psi)> =
           -\frac{\beta(\beta+4)}{\beta+6} \nonumber \\
       C_4(g)&\equiv&<\frac{1}{N^3}\tr(\pbs\psi\pbs\psi)> =
           \frac{\beta^2 + 5\beta + 6}{g(\beta + 6)}.
\eeqs
The result for $C_4$ is rather curious and applies only to fermionic matrix
models; for all other models a disconnected term $\approx C_2^2$ would survive
after tracing.

Finally, by a decomposition similar to~(\ref{clustermatr}), we see that the
vertex functions are again related to $C_2$ and $C_4$  as in
appendix~\ref{appendixvect}:
\beq
      \Gamma_2(g) = C_2(g)^{-1} \quad\mbox{and}\quad
      \Gamma_4(g) = C_4(g) C_2(g)^{-4}.
\eeq
These are the quantities used in section~\ref{appmatrix} to obtain the
renormalization group equations.

\end{document}